# The Baryon Fraction and Velocity–Temperature Relation in Galaxy Clusters : Models versus Observations

Lori M. Lubin, Renyue Cen, Neta A. Bahcall & Jeremiah P. Ostriker
Princeton University Observatory
Peyton Hall
Princeton, NJ 08544-0001

## ABSTRACT

The observed baryon fraction and velocity–temperature relation in clusters of galaxies are compared with hydrodynamic simulations in two cosmological models : standard ($\Omega = 1$) and a low-density flat ($\Omega = 0.45$ and $\lambda = 0.55$) CDM models, normalized to the COBE background fluctuations. The observed properties of clusters include the velocity dispersion versus temperature relation, the gas mass versus total mass relation, and the gas mass fraction versus velocity dispersion relation. We find that, while both cosmological models reproduce well the *shape* of these observed functions, only low-density CDM can reproduce the observed *amplitudes*. We find that $\sigma \sim T^{0.5 \pm 0.1}$, as expected for approximate hydrostatic equilibrium with the cluster potential, and the ratio of gas to total mass in clusters is approximately constant for both models. The *amplitude* of the relations, however, differs significantly between the two models. The low-density CDM model reproduces well the *average* observed relation of $M_{gas} = (0.13 \pm 0.02)\, M\, h_{50}^{-1.5}$ for clusters, while $\Omega = 1$ CDM yields a gas mass that is *three* times lower ($M_{gas} = 0.045 \pm 0.004\, M\, h_{50}^{-1.5}$) with both gas and total mass measured within a fiducial radius of 1.5 $h^{-1}$ $Mpc$. The cluster gas mass fraction reflects approximately the baryon fraction in the models, $\Omega_b/\Omega$, with a slight anti-bias. Therefore, due to the low baryon density given by nucleosynthesis, $\Omega_b \simeq 0.06\, h_{50}^{-2}$, $\Omega = 1$ models produce too few baryons in clusters compared with observations. Scaling our results as a function of $\Omega$, we find that a low-density CDM model, with $\Omega \sim 0.3 - 0.4$, best reproduces the observed mean baryon fraction in clusters. The observed $\beta$ parameter of clusters, $\beta = \sigma^2/(kT/\mu m_p) = 0.94 \pm 0.08$ discriminates less well between the models; it is consistent with that produced by low-density CDM ($1.10 \pm 0.22$), while it is slightly larger than expected but still consistent with $\Omega = 1$ ($0.70 \pm 0.14$).

## 1. Introduction

Clusters of galaxies provide a powerful probe of cosmology. As observations of optical and X-ray properties of clusters improve, comparisons of the observed properties with expectations from different cosmological models can place strong constraints on theories for the origin of structure in the universe (e.g., Edge et al. 1990; Bahcall & Cen 1992; White et al. 1993b; Evrard et al. 1993; Ostriker 1993; Kang et al. 1994; Bryan et al. 1994; Cen & Ostriker 1994). It is well known (White et al. 1993b) that the high gas mass or baryon fraction observed in clusters, $\Omega_b/\Omega \sim 0.15\ h^{-1.5}$, poses a problem for an $\Omega = 1$ universe when combined with the low value of the baryon density given by primordial nucleosynthesis, $\Omega_b \sim 0.06\ h_{50}^{-2}$ (Walker et al. 1991). Possible resolution of this conflict include : 1) a low-density universe; 2) a baryon density $\Omega_b$ that is larger than expected from nucleosynthesis arguments; 3) segregation of baryons and dark matter, with a significant overdensity of baryons in clusters of galaxies. In this paper, we test points (1) and (3) above by studying two specific cosmological models, an $\Omega = 1$ and a flat $\Omega = 0.45$ cold dark matter model (SCDM and LCDM, respectively) and compare their predictions with observations of the gas mass and baryon fraction in clusters and the velocity-temperature relation. We use a new hydrodynamic plus dark matter code to simulate the evolution of structure in these models. Both models assume the baryon density value given by nucleosynthesis and are normalized to the observed COBE fluctuations of the microwave background radiation (Smoot et al. 1992).

The SCDM model has known difficulties; when normalized to COBE observations it cannot reproduce some basic observations such as the cluster mass and correlation functions, the galaxy pairwise velocities on small scales, and some X-ray properties of clusters (Davis et al. 1985; Maddox et al. 1990; Cen & Ostriker 1992; Bahcall & Cen 1992; Ostriker 1993; Kang et al. 1994; Peebles 1994 and references therein). A low-density CDM model, on the other hand, eliminates most of these inconsistencies and provides a good match to the data (Efstathiou 1992; Bahcall & Cen 1992; White et al. 1993a; Kofman et al. 1993; Cen & Ostriker 1994; Peebles 1994). We use the observed correlations of optical and X-ray cluster properties to further test the applicability of the low-density CDM model. We examine, from observations and simulations, the relations between the cluster velocity dispersion ($\sigma$), temperature ($T$) of the hot intracluster medium, the amount of gas mass ($M_{gas}$), and total virial mass ($M$) within a given radius of the cluster center. Observationally, a strong correlation exists between $\sigma$ and $T$ which is consistent with that expected from the isothermal, hydrostatic model (Mushotzky 1984; Sarazin 1988; Edge & Stewart 1991; Lubin & Bahcall 1993; Bird et al. 1995). The $\beta$ parameter, which relates the average energy per unit mass in the galaxies to that in the gas, indicates that the galaxies and the intracluster gas trace the same potential; that is, $\beta \simeq 1$ (Edge & Stewart 1991; Lubin & Bahcall 1993). The amount of gas in clusters is also highly correlated with $\sigma$ and the total mass of the cluster (Jones & Forman 1992) and reflects the baryon fraction in clusters. We compare these observed relations with the simulated clusters, finding that the low-density CDM model provides a good fit to the observed cluster properties, while the standard CDM model cannot reproduce the observed relations.

We describe the observations in Sect. 2, the model simulations in Sect. 3, and present a comparison between the two in Sect. 4. We summarize our conclusions in Sect. 5.

## 2. Observations

Recent X-ray and optical observations of clusters of galaxies provide a large sample of rich clusters with measured galaxy velocity dispersions (e.g. Struble & Rood 1991; Zabludoff 1993), temperatures of the hot intracluster gas (Edge et al. 1990; Edge & Stewart 1991; David et al. 1993), and the amount of gas mass in the cluster (e.g. Arnaud et al. 1992; Jones & Forman 1992; Böhringer 1994), all within a radius of typically $R_A \simeq 1.5\ h^{-1}\ Mpc$. (We use the notation $H_o = 100\ h\ km\ s^{-1}\ Mpc^{-1}$, with $h_{50}$ indicating $H_o = 50$.) Estimates of the cluster's total virial mass within this radius can also be calculated using both galaxy velocities and the X-ray temperature (e.g. Hughes 1989; Bahcall & Cen 1992; Henry et al. 1993 and references therein). For the best observed clusters the optical and X-ray cluster mass determinations yield consistent results for mass within a given radius. The galaxy velocities and the gas temperatures are consistent with each other (that is, $\sigma^2 \sim kT/\mu m_p$, where $\sigma$ is the line-of-sight velocity dispersion of the galaxies in the cluster; Lubin & Bahcall 1993), and the gas and galaxy distributions in the clusters are also generally similar (Bahcall & Lubin 1994; Gerbal et al. 1994) for this sample.

In this study, we use all rich clusters which have velocity dispersions with more than twenty measured galaxy redshifts per cluster (Struble & Rood 1991), measured X-ray temperatures (David et al. 1993), and gas masses (Arnaud et al. 1992; Jones & Forman 1992; Böhringer 1994), all within $R_A \simeq 1.5\ h^{-1}\ Mpc$ of the cluster center (except for the temperature which is typically measured within half that radius). The sample corresponds to 41 rich clusters for which $\sigma$ and $T$ are available and 26 clusters with $\sigma$ and $M_{gas}$. The mass of each cluster is determined from the virial theorem assuming an isothermal distribution within $1.5\ h^{-1}\ Mpc$ : $M(R \leq 1.5\ h^{-1}\ Mpc) = \frac{2\sigma^2 R}{G}$. X-ray observations of the clusters indicate that the gas mass in rich clusters (within $1.5\ h^{-1}\ Mpc$) ranges from $\sim 10^{13}$ to $10^{14}\ h^{-2.5}\ M_\odot$. The fraction of the total cluster mass that exists in hot gas ranges from $M_{gas}/M \simeq (0.02\ to\ 0.13)\ h^{-1.5}$ (Arnaud et al. 1992; Jones & Forman 1992; Böhringer 1994;

Elbaz, Arnaud, & Böhringer 1995).

For comparison, we also include four X-ray groups of galaxies : two small groups (the NGC 2300 group and Hickson Compact Group 62) observed by the ROSAT satellite (Mulchaey et al. 1993; Ponman & Bertram 1993) and two Morgan poor clusters (MKW4 and AWM4; Beers et al. 1984; Malumuth & Kriss 1986). These groups have X-ray temperatures in the range of $1 - 4$ $keV$ and apparent gas mass ratios of $M_{gas}/M \simeq 0.01 - 0.04$ $h^{-1.5}$, within the measured radii of $\sim 0.2 - 0.5$ $h^{-1}$ $Mpc$ (above references).

**Velocity–Temperature Relation.** The observed $\sigma - T$ relation for clusters is presented in Figure 1 (see also Lubin & Bahcall 1993). There exists a clear correlation between the two parameters with $\sigma \sim T^{0.5}$, as expected from a hydrostatic cluster approximation. The best-fit relation (determined from a weighted $\chi^2$ fitting) yields $\sigma = (332 \pm 52) (kT)^{0.6 \pm 0.1}$ km s$^{-1}$ (where $kT$ is in keV). The observed relation indicates that there is approximately equal energy per unit mass in the galaxies and the gas. The best-fit $\beta$ parameter

$$\beta = \sigma^2/(kT/\mu m_p), \qquad (1)$$

where $\mu m_p$ is the average particle mass, is $0.94 \pm 0.08$ and $\beta$(median) $= 0.98$ (see Figure 1 and Lubin & Bahcall 1993). Groups of galaxies are consistent with the above relation, extending it to lower $\sigma$ and $T$ values.

**Gas Mass.** The relation between the observed gas mass and the total virial mass of the clusters, within $R_A \simeq 1.5$ $h^{-1}$ $Mpc$, is presented in Figure 2. The error-bars represent typical $1\sigma$ uncertainties in $M_{gas}$ ($10^{\pm 0.12}$) and total mass $M$ ($10^{\pm 0.2}$), obtained from the above references. While the scatter is large (mostly due to large uncertainties in the mass which include the rms scatter in the $\sigma - T$ relation), a correlation between the parameters is evident with an unweighted $\chi^2$ fitting (including the four groups of galaxies; see above) of $M_{gas} h^{2.5} \simeq 2.3 \times 10^{-4} (M h)^{1.16 \pm 0.12}$. The best-fit linear relation satisfies $M_{gas} = (0.047 \pm 0.007) M h^{-1.5} [\simeq (0.13 \pm 0.02) M h_{50}^{-1.5}]$; the error-bar reflects the uncertainty of the best-fit average value. The dependence of the observed gas mass fraction in clusters, $M_{gas}/M$, on the cluster velocity dispersion is presented in Figure 3. The gas fraction ranges from $\sim 0.02$ to $0.13$ $h^{-1.5}$ for the clusters and is essentially independent of $\sigma$. The gas fraction provides a lower limit to the baryon fraction in the clusters. (Note that Figures 2 and 3 are presented for the lower $h$ values appropriate to the cluster simulations; see Sect. 4). A comparison of the above relations with expectations from model simulations is discussed in Sect. 4.

## 3. Model Simulations

A three-dimensional, shock-capturing, hydrodynamic code (Ryu et al. 1993) is utilized to determine the distribution of hot gas in clusters in two CDM model universes. The simulations have a box size of $L = 85\ h^{-1}\ Mpc$ with $N = 270^3$ cells and $135^3$ dark matter particles, with a corresponding resolution of $\sim 0.31\ h^{-1}\ Mpc$. A detailed description of the simulations and the properties of the X-ray clusters in the two models were presented in Kang et al. (1994) and Cen & Ostriker (1994). The two cosmological models are: (1) the standard CDM model (SCDM) with $\Omega = 1$, $h = 0.5$, $\Omega_b = 0.06$, and $\sigma_8 = 1.05$; and (2) a flat low-density CDM model (LCDM) with $\Omega = 0.45$, $\lambda = 0.55$, $h = 0.6$, $\Omega_b = 0.042$, and $\sigma_8 = 0.77$. Both models are normalized to the first year COBE background fluctuations on large scales (Smoot et al. 1992); $\sigma_8$ describes the rms mass fluctuations on a top-hat sphere of radius $8\ h^{-1}\ Mpc$. The baryon density, $\Omega_b$, is consistent with expectations from light element nucleosynthesis (Walker et al. 1991).

The simulated X-ray clusters are identified by calculating the total X-ray luminosity due to thermal Bremsstrahlung for each cell given the cell density and temperature. Cells with total X-ray luminosities higher than $10^{38}\ ergs\ s^{-1}$ are considered "X-ray bright" cells. Local maxima are found by comparing the X-ray luminosity of each X-ray bright cell with that of 124 neighboring cells and are identified as the centers of the X-ray clusters (see also Kang et al. 1994). We examine the velocity dispersion of the dark matter particles in the cluster, the total mass, and the gas mass of the clusters, within $1.5\ h^{-1}\ Mpc$ radius of each cluster center; the emission-weighted temperature of the intracluster gas is computed within $0.75\ h^{-1}\ Mpc$ radius of each cluster center, consistent with the observations. The volume studied includes approximately 450 cells so numerical resolution effects should not be serious. We discuss the corrections below. A comparison between the observations and the two cosmological simulations is presented in the following section.

**Resolution Calibrations.** The sensitivity of the results to the limited resolution in our hydrodynamic simulations ($0.31\ h^{-1}\ Mpc$) was studied and corrected as follows. First, we ran two pure N-body simulations for the SCDM model with identical initial conditions and the same box size, $L = 128\ h^{-1}\ Mpc$. One of the two simulations has the same numerical resolution and the same force computing scheme PM+FFT as our hydrodynamic simulations ($0.31\ h^{-1}\ Mpc$); the other simulation has a higher resolution ($0.025\ h^{-1}\ Mpc$) based on the $P^3M$ scheme but utilizing a special computer chip (GRAPE) to solve the PP part of the force computation (Brieu, Summers & Ostriker 1995). Since the resolution of the $P^3M$ simulations (Summers, Cen & Ostriker 1995) is smaller than the core size of clusters, it can be treated as having sufficient resolution for the purpose of calibrating our low resolution results. Because the two simulations have identical initial conditions, we are able to identify each cluster in one simulation with its counterpart in the other (for details, see Summers, Cen & Ostriker 1995). Having made such a one-to-one correspondance, we can compute the ratios of the high resolution (HR) to the low resolution (LR) quantities $M_{HR}/M_{LR}$ and $\sigma_{HR}/\sigma_{LR}$ as a function of the cluster mass $M_{LR}$. This allows us to make corrections to $M$ and $\sigma$ in the hydrodynamic simulation of the SCDM model presented here. The above calibration procedure is repeated for the LCDM model. We then determine and apply the following corrections (generally $\leq 30\%$) to the dark matter velocity dispersion ($\sigma_{DM}$) and the cluster mass ($M_{15}$; in units of $10^{15}$ $M_\odot$) : $\sigma_{HR}/\sigma_{LR} = 1.110 - 0.158$ $\mathrm{Log}_{10}$ $(M_{15}\ h_{50}^{-1})$ and $M_{HR}/M_{LR} = 1.083 + 0.109$ $\mathrm{Log}_{10}$ $(M_{15}\ h_{50}^{-1})$ for SCDM; and $\sigma_{HR}/\sigma_{LR} = 1.046 - 0.174$ $\mathrm{Log}_{10}$ $(M_{15}\ h_{60}^{-1})$ and $M_{HR}/M_{LR} = 1.103 + 0.065$ $\mathrm{Log}_{10}$ $(M_{15}\ h_{60}^{-1})$ for LCDM.

Second, we made two hydrodynamic simulations of a power law model $P_k = k^{-1}$ with box sizes of $L = 90$ and $45$ $h^{-1}$ $Mpc$, respectively. Both models have identical parameters $\sigma_8 = 1.0$, $\Omega = 1$, and $\Omega_b = 0.06$, and the same number of cells ($288^3$) and dark matter particles ($144^3$). The X-ray clusters are identified in the same way for both simulations. Since the two simulations have different sizes, one-to-one mapping of clusters in the two simulations is not possible. Therefore, our calibration is based on statistical averaging. We find that, at a fixed cluster mass, the average temperature of clusters in one simulation is equal to the average temperature of clusters in the other simulation. On the other hand, the differences in $M$ and $\sigma$ between the two simulations are comparable to those found between the low resolution PM simulations and the very high resolution P$^3$M simulations of the first test discussed above, indicating the $L = 45$ $h^{-1}$ $Mpc$ hydrodynamic simulation can be considered to have sufficiently high resolution for the purpose of temperature calibration.

The fact that $\sigma$ increases slightly with increasing resolution while our computed values of $T$ are independent of resolution may be due to the fact that $\sigma$ is calculated within $r < 1.5$ $h^{-1}$ $Mpc$, and $T$ is the emission measured temperature (weighted toward the central cluster regions by a density–square term) within $r < 0.75$ $h^{-1}$ $Mpc$, where the temperature is approximately constant.

## 4. Comparison of Observations and Model Simulations

### 4.1. Velocity–Temperature Relation

The observed $\sigma - T$ relation (Sect. 2) is compared with the results of the SCDM and LCDM (including corrections) models in Figure 1. We include all simulated clusters with X-ray temperatures greater than 1 keV. Because the simulations contain no galaxies, the simulated clusters yield a relation between the dark matter velocity dispersion ($\sigma_{DM}$) and the X-ray temperature ($T_{sim}$).

Both cosmological scenarios produce a $\sigma_{DM} - T_{sim}$ relation that has a *shape* similar to the observed shape of the relation; this shape is consistent with the gas being in hydrostatic equilibrium with the binding cluster potential, i.e. $\sigma_{DM} \propto T_{sim}^{0.5\pm0.1}$. The model simulations differ, however, in their respective $\beta$ values (as defined in Eq. 1), i.e., in the amplitude of the $\sigma - T$ relation. The best $\chi^2$ fit of the $\sigma_{DM} - T_{sim}$ simulated relation yields $\beta_{DM}$ of $1.10 \pm 0.02$ for SCDM (consistent with Navarro et al. 1995) and $1.36 \pm 0.03$ for LCDM. Because the dark matter velocity dispersion is used rather than the galaxy velocity dispersion ($\sigma_{gal}$), we cannot directly compare the observed $\beta$ with $\beta_{DM}$; however, simulations which include galaxies (e.g. Carlberg & Dubinski 1991; Couchman & Carlberg 1992; Cen & Ostriker 1992) suggest a small velocity bias within clusters : $b_v \equiv \sigma_{gal}/\sigma_{DM} \simeq 0.8$ for $\Omega = 1$ CDM and $b_v \simeq 0.9$ for low-density CDM. This implies that the galaxy $\beta$ for the simulations is $\beta_{sim} = \beta_{DM} \times b_v^2 \simeq 0.70 \pm 0.14$ for SCDM and $1.10 \pm 0.22$ for LCDM, assuming a 10% uncertainty in $b_v$ (see Table 1). The observed value, $\beta = 0.94 \pm 0.08$, is consistent with that produced by low-density CDM. The standard $\Omega = 1$ CDM cosmology, with $\beta_{sim} = 0.70 \pm 0.14$, yields a slightly low $\beta$ value in comparison with the observations but is still consistent within $2\sigma$ (mostly due to the large uncertainty in $b_v^2$) [see Figure 1; both $\beta_{DM}$ and the inferred galaxy $\beta_{sim}$ are presented for comparison with the observations.]

The SCDM model produces many more rich (high $T_{sim}$) clusters than the LCDM model (Figure 1). This is due to the fact that an SCDM model with $\sigma_8 \sim 1$ overproduces the most massive clusters relative to the observed cluster mass function and relative to LCDM (Bahcall & Cen 1992), while the LCDM model produces a cluster frequency that is consistent with observations. Since rich clusters are rare, and the simulation box size is not very large, only a small number of the richest clusters are seen in the low-density CDM model.

### 4.2. Gas Mass in Clusters

The relation between the gas mass and the total mass of the simulated clusters is compared with observations in Figure 2. The amount is defined to be that within a fiducial radius of $1.5h^{-1}$ $Mpc$. The observed and simulated data are plotted for the relevant $H_o$ of the simulation ($h = 0.5$ for SCDM and $h = 0.6$ for LCDM). The shape of the simulated $M_{gas} - M$ relation is consistent with observations; it follows, approximately, $M_{gas} \propto M$. However, the amplitude of the relation, or equivalently the gas mass fraction in the clusters, $M_{gas}/M$, which represents a lower limit to the baryon fraction in clusters, differs significantly in the two models. The SCDM model yields $M_{gas} = 0.045 \pm 0.004$ $M$ for $h = 0.5$, i.e. lower gas masses by a factor of $\sim 3$ for a given cluster mass than the observed best-fit relation $M_{gas}(\text{obs}) = 0.13 \pm 0.02$ $M$ $h_{50}^{-1.5}$. LCDM, on the other hand, yields $M_{gas} = 0.077 \pm 0.005\ M$ for $h = 0.6$, more consistent with the data (see Table 1). An even lower mass density of $\Omega \sim 0.35$ will provide a better match with the data, yielding $M_{gas}/M = 0.10 \pm 0.01$ ($h = 0.6$) [Table 1].

It is interesting to compare the average global baryon to total mass ratios $(\Omega_b/\Omega) = (0.060, 0.093)$ to the cluster gas to total mass ratios $(M_{gas}/M) = (0.045 \pm 0.004, 0.077 \pm 0.005)$ of the two models (SCDM, LCDM), respectively. We note that the gas is relatively underrepresented in the clusters (i.e. "anti-biased") by $17 - 25\%$. We believe that this effect is real as it has been observed in all simulations performed to date (Cen & Ostriker 1992, 1993; Katz, Hernquist, & Weinberg 1992; Evrard, Summers, & Davis 1994; Kang et al. 1994; Bryan et al. 1994). It reflects, presumably, an hydrodynamic effect whereby some small fraction of the shocked gas is ejected from the central parts of the clusters. This effect exacerbates the difficulties made for $\Omega = 1$ models by the cluster gas mass observations. That is, the observed, high gas to total mass ratio in clusters in fact must slightly underestimate the true global baryon to total mass ratio.

The relation between the gas mass fraction and the line-of-sight velocity dispersion in the simulated clusters is compared with observations in Figure 3. The simulated results present the dark matter velocity dispersion ($\sigma_{DM}$) of the clusters. The simulated galaxy velocity dispersion would be lower by the appropriate bias factor (see Sect. 4.1). The clusters include all systems with $T_{sim} > 1\ keV$. The figure is presented for the relevant $H_o$ of the individual simulation. Neither the observations nor the simulations show any significant dependence of $M_{gas}/M$ on $\sigma$. The best-fit $M_{gas}/M$ ratios are presented in Table 1. As expected from Figure 2, the SCDM model produces rich clusters with a gas mass ratio which is too small by a factor of three compared to the observations. The LCDM model reproduces the observations much better (though the scatter in the observations is much larger than the scatter in the simulations). This trend is consistent with expectations given the low baryon density of nucleosynthesis (Sect. 1).

The constant $M_{gas}/M$ ratio in the simulations, independent of $\sigma_{DM}$ or $T_{sim}$, implies that the total gas mass is proportional to the cluster potential, i.e. $M_{gas} \propto M \propto \sigma_{DM}^2$, within a given radius.

## 5. Conclusions

We compare observed optical and X-ray properties of clusters of galaxies with hydrodynamic simulations of clusters in two cosmological models : standard ($\Omega = 1$) CDM and low-density CDM models, normalized to the COBE background fluctuations. The observed properties include the velocity dispersion – temperature relation ($\sigma - T$), the gas mass versus total mass relation ($M_{gas} - M$), and the gas mass fraction versus velocity dispersion relation ($M_{gas}/M - \sigma$) for clusters. Our principal conclusions are summarized below.

1. We find that both a low-density CDM model and an $\Omega = 1$ model are reasonably consistent with the observed optical – X-ray relation $\sigma - T$ of galaxy clusters.

2. The low-density CDM model yields a $M_{gas} - M$ relation that is consistent with observations, with $M_{gas} = (0.077 \pm 0.005) \, M$ (for $h = 0.6$ and within $R_A \sim 1.5 \, h^{-1} \, Mpc$ of the cluster center), as compared with the observed mean of $M_{gas} = (0.10 \pm 0.02) \, M \, h_{60}^{-1.5}$. (The agreement improves for $\Omega \sim 0.35$). Standard CDM, however, exhibits lower gas mass ratios than observed by a factor of $\sim 3$ with $M_{gas} = (0.045 \pm 0.004) \, M$ ($h = 0.5$), as compared with the observed mean of $M_{gas} = 0.13 \pm 0.02 \, M \, h_{50}^{-1.5}$ (Table 1; Figures 2–3). This result is expected, given the low baryon density of $\Omega_b \simeq 0.06 \, h_{50}^{-2}$ implied by nucleosynthesis. Neither the simulations nor the observations show a significant dependence of the ratio $M_{gas}/M_{tot}$ on increasing $\sigma$ (or $M$) for the rich clusters considered in this paper.

3. The cluster gas mass fractions in the two simulations reflect the baryon fraction in the models, but with a small ($\sim 20\%$) antibias. That is, in the simulations, gas is slightly underrepresented within $1.5 \, h^{-1} Mpc$ of cluster centers. This effect quantitatively strenghtens the conclutions of White et al. (1993b).

We are grateful to Frank Summers for allowing us to use the high resolution GRAPE P³M simulations. It is a pleasure to acknowledge the help of NCSA and PSC for allowing us to use their Convex-3880 and Cray C90 supercomputers. This research is supported in part by NASA grants NAGW-2448 and NGT-51295, NSF grants AST91-08103 and AST93-15368, and NSF HPCC grant ASC-9318185. RC and JPO would like to thank the hospitality of ITP during their stay when this work was completed, and the financial support from ITP through the NSF grant PHY94-07194.

– 11 –

## Figure Captions

Figure 1. Observed and simulated one-dimensional cluster velocity dispersion versus the intracluster gas temperature for clusters and groups of galaxies. The best-fit $\beta$ parameter for the observed 41 rich clusters and the simulated clusters (both for the dark-matter, $\beta_{DM}$, and the galaxies, $\beta_{sim}$) are shown by the best-fit lines. (a) standard $\Omega = 1$ CDM; (b) $\Omega = 0.45$ CDM.

Figure 2. The observed and simulated gas mass ($M_{gas}$) versus total mass ($M$) for rich clusters (within $R_A \simeq 1.5\ h^{-1}\ Mpc$) and groups of galaxies (shown at their observed radii; Sect. 2). The error-bars represent typical $1\sigma$ uncertainties in the observed masses. (a) standard $\Omega = 1$ CDM ($h = 0.5$); (b) $\Omega = 0.45$ CDM ($h = 0.6$).

Figure 3. The observed and simulated gas mass fraction ($M_{gas}/M$) versus line-of-sight velocity dispersion ($\sigma$) for rich clusters and groups of galaxies. A typical $1\sigma$ uncertainty is shown in the right of each panel. The simulated results (open squares) present the dark-matter velocity dispersion of the clusters; the galaxy velocity dispersion is lower by $b_v \sim 0.8$ for SCDM and $b_v \sim 0.9$ for LCDM. (a) SCDM ($h = 0.5$); (b) LCDM ($h = 0.6$).

Table 1 : Observed versus Simulated Cluster Properties

| Model | $\beta$ | | $< M_{gas}/M >$ | |
|---|---|---|---|---|
| | Observed | Simulated[a] | Observed | Simulated |
| SCDM $\Omega = 1$ | $0.94 \pm 0.08$ | $0.70 \pm 0.14$ | $0.13 \pm 0.02$ ($h = 0.5$) | $0.045 \pm 0.004$ ($h = 0.5$) |
| LCDM $\Omega = 0.45$ | $0.94 \pm 0.08$ | $1.10 \pm 0.22$ | $0.10 \pm 0.02$ ($h = 0.6$) | $0.077 \pm 0.005$[b] ($h = 0.6$) |

[a] The simulated $\beta$ is $\beta_{sim} = \sigma^2_{gal}/(kT_{sim}/\mu m_p)$ where $\sigma_{gal}$ is estimated from $\sigma_{DM}$ and the appropriate velocity bias (see Sect. 4.1).

[b] The LCDM gas mass fraction increases to $0.10 \pm 0.01$ ($h = 0.6$) for $\Omega = 0.35$.

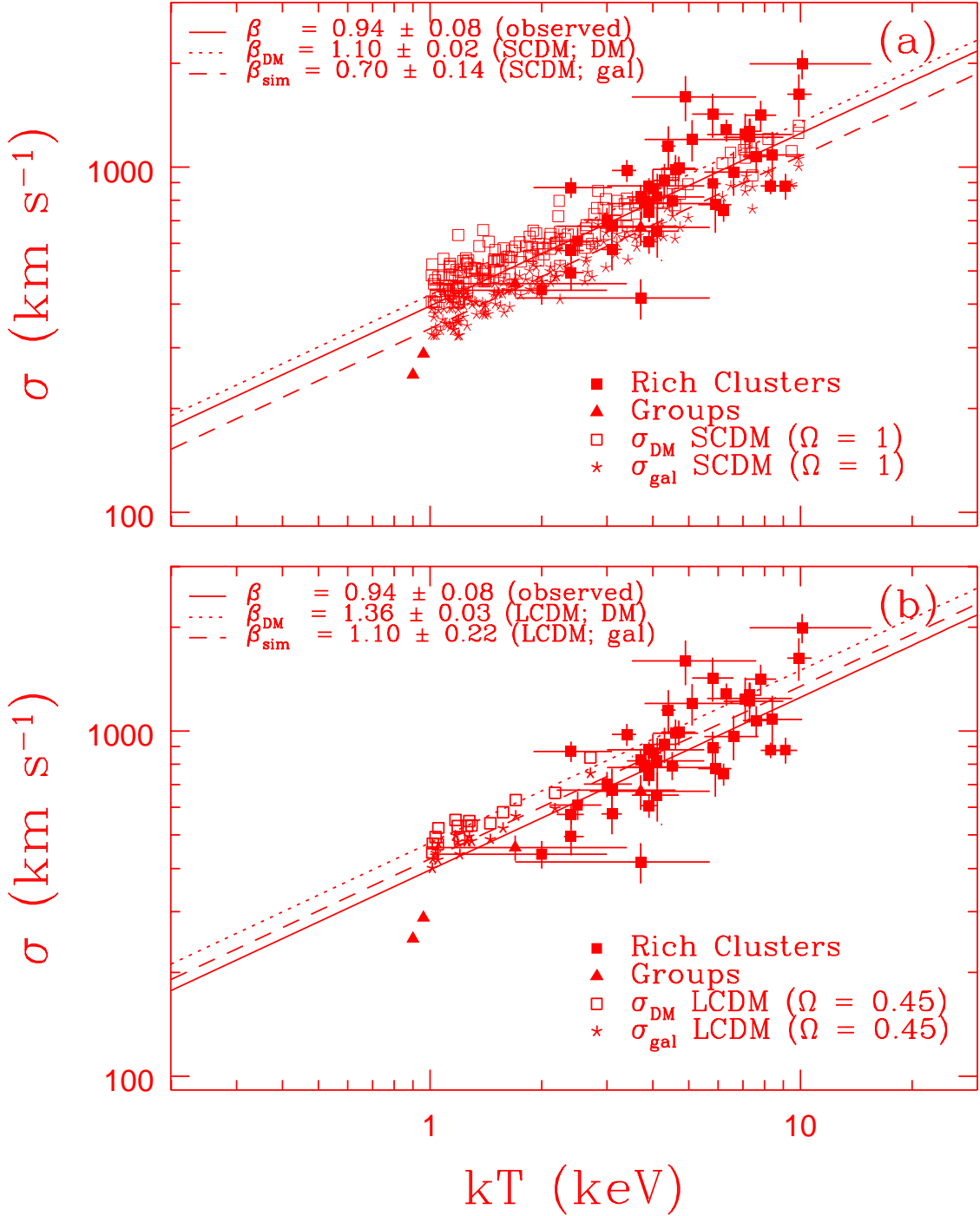

Fig. 1.— Observed and simulated one-dimensional cluster velocity dispersion versus the intracluster gas temperature for clusters and groups of galaxies. The best-fit $\beta$ parameter for the observed 41 rich clusters and the simulated clusters (both for the dark-matter, $\beta_{DM}$, and the galaxies, $\beta_{sim}$) are shown by the best-fit lines. (a) standard $\Omega = 1$ CDM; (b) $\Omega = 0.45$ CDM.

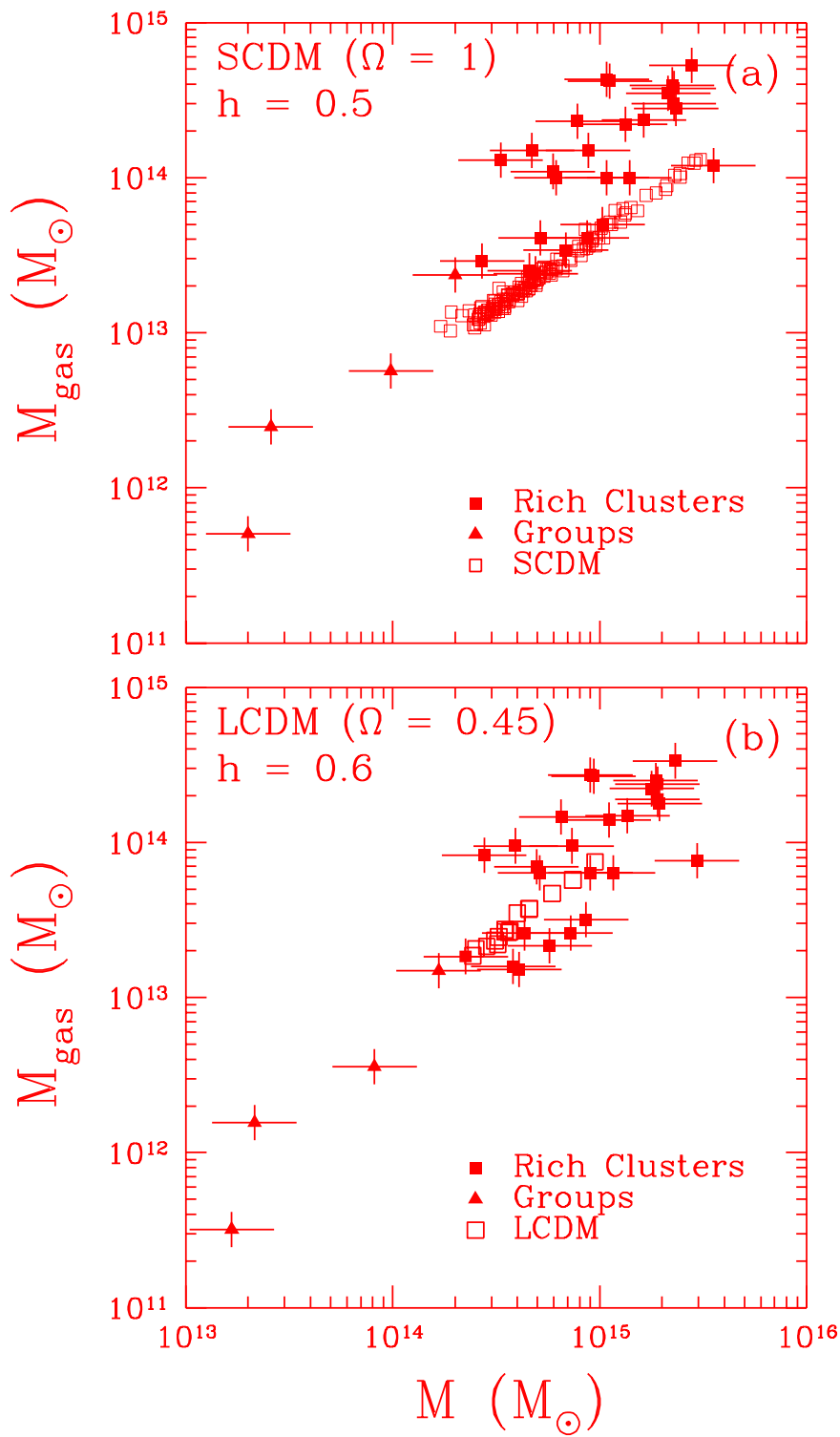

Fig. 2.— The observed and simulated gas mass ($M_{gas}$) versus total mass ($M$) for rich clusters (within $R_A \simeq 1.5\ h^{-1}\ Mpc$) and groups of galaxies (shown at their observed radii; Sect. 2). The error-bars represent typical $1\sigma$ uncertainties in the observed masses. (a) standard $\Omega = 1$ CDM ($h = 0.5$); (b) $\Omega = 0.45$ CDM ($h = 0.6$).

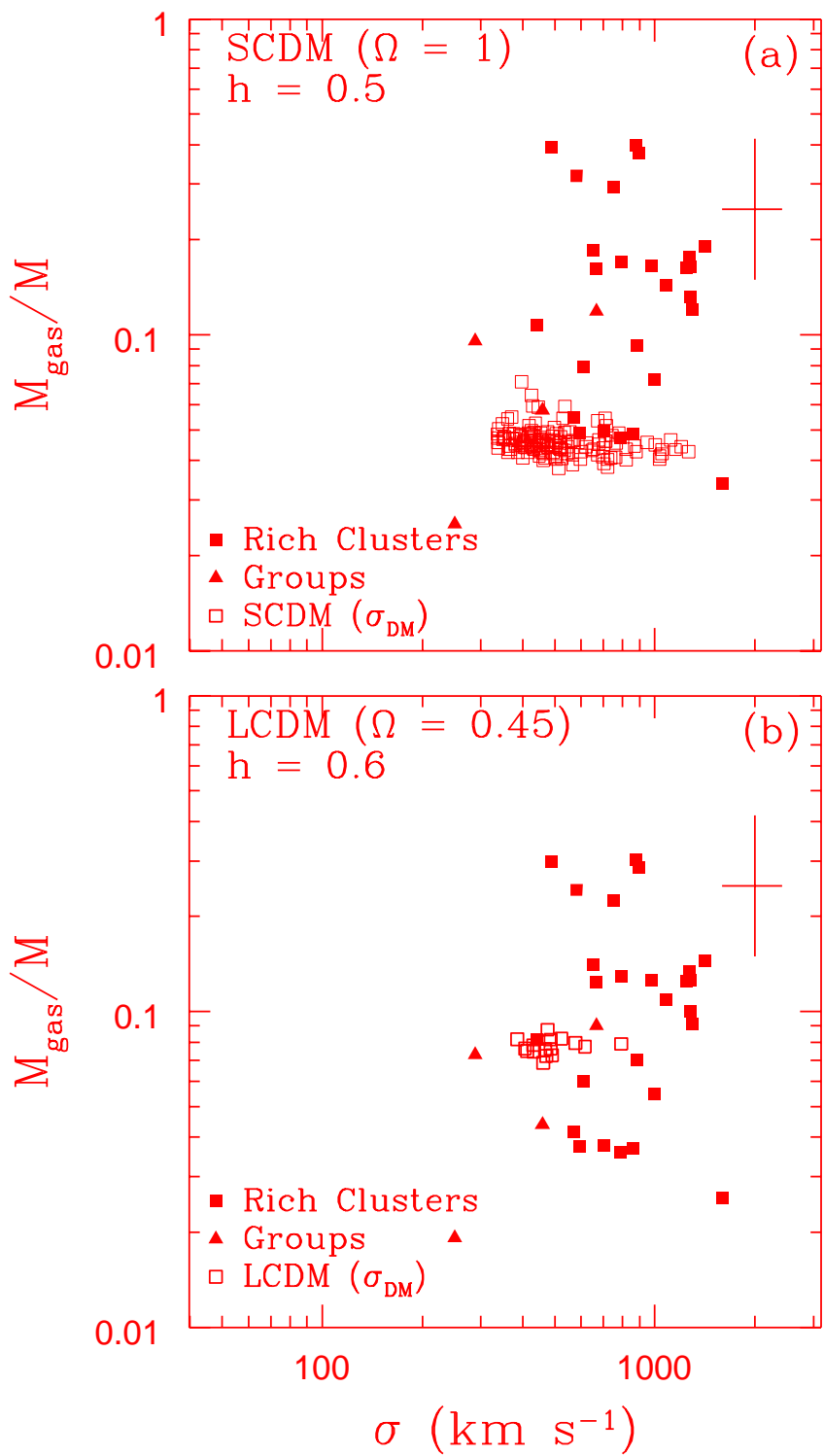

Fig. 3.— The observed and simulated gas mass fraction ($M_{gas}/M$) versus line-of-sight velocity dispersion ($\sigma$) for rich clusters and groups of galaxies. A typical $1\sigma$ uncertainty is shown in the right of each panel. The simulated results (open squares) present the dark-matter velocity dispersion of the clusters; the galaxy velocity dispersion is lower by $b_v \sim 0.8$ for SCDM and $b_v \sim 0.9$ for LCDM. (a) SCDM ($h = 0.5$); (b) LCDM ($h = 0.6$).